\documentclass[10pt,journal,final]{IEEEtran}
\makeatletter
\renewcommand\normalsize{%
\@setfontsize\normalsize\@xpt\@xiipt
\abovedisplayskip 4\p@ \@plus2\p@ \@minus5\p@
\abovedisplayshortskip \z@ \@plus3\p@
\belowdisplayshortskip 6\p@ \@plus3\p@ \@minus3\p@
\belowdisplayskip \abovedisplayskip
\let\@listi\@listI}
\makeatother

\usepackage{float}
\usepackage{color}
\usepackage{graphicx}
\graphicspath{{./Figures/}}
\usepackage{amsmath}
\usepackage[justification=centering]{caption}

\usepackage{verbatim}

\usepackage{multirow}
\usepackage{makecell}

\usepackage{amsmath}

\usepackage{amssymb}
\usepackage{algorithmic}
\usepackage{algorithm}
\usepackage{booktabs}
\usepackage{cite}
\usepackage[colorlinks,linkcolor=blue,citecolor=blue,urlcolor=black]{hyperref}

\begin{document}
\title{Optimal Scheduling of Isolated Microgrids Using Automated Reinforcement Learning-based Multi-period Forecasting}
\author{Yang~Li,~\IEEEmembership{Senior Member,~IEEE,}
       Ruinong~Wang, Zhen~Yang
\thanks{This work is partly supported by the Natural Science Foundation of Jilin Province, China under Grant No. 2020122349JC. (Corresponding author: Yang Li.)
\par Y. Li and R. Wang are with the School of Electrical Engineering, Northeast Electric Power University, Jilin 132012, China (e-mail: liyang@neepu.edu.cn; 624895223@qq.com).
\par Z. Yang is with the State Grid Beijing Electric Power Company, Xicheng District, Beijing 100032, China (e-mail: 1678084931@qq.com).}}

\markboth{IEEE Transactions on Sustainable Energy}%
{Shell \MakeLowercase{\textit{et al.}}: Bare Demo of IEEEtran.cls for IEEE Journals}

\maketitle
\begin{abstract}
    In order to reduce the negative impact of the uncertainty of load and renewable energies outputs on microgrid operation, an optimal scheduling model is proposed for isolated microgrids by using automated reinforcement learning-based multi-period forecasting of renewable power generations and loads. Firstly, a prioritized experience replay automated reinforcement learning (PER-AutoRL) is designed to simplify the deployment of deep reinforcement learning (DRL)-based forecasting model in a customized manner, the single-step multi-period forecasting method based on PER-AutoRL is proposed for the first time to address the error accumulation issue suffered by existing multi-step forecasting methods, then the prediction values obtained by the proposed forecasting method are revised via the error distribution to improve the prediction accuracy; secondly, a scheduling model considering demand response is constructed to minimize the total microgrid operating costs, where the revised forecasting values are used as the dispatch basis, and a spinning reserve chance constraint is set according to the error distribution; finally, by transforming the original scheduling model into a readily solvable mixed integer linear programming via the sequence operation theory (SOT), the transformed model is solved by using CPLEX solver. The simulation results show that compared with the traditional scheduling model without forecasting, this approach manages to significantly reduce the system operating costs by improving the prediction accuracy.

\end{abstract}
    
\begin{IEEEkeywords}
    Microgrid, optimal scheduling, automated reinforcement learning, uncertainty handling, single-step multi-period forecasting, sequence operation theory.      
\end{IEEEkeywords}

\section{Introduction}\label{sec.intro}
\IEEEPARstart{W}{ith} the gradual depletion of fossil fuels, the deteriorating ecological environment, combined with traditional centralized power supply exposes many shortcomings, the proportion of renewable power generation in power system is increasing due to its sustainability and environmental friendliness \cite{zargar2019development}. As an effective carrier of distributed generations (DGs), a microgrid has been widely used in power systems in recent years. In-depth research and accelerated construction of microgrids can promote the large-scale integration of DGs and renewable energy sources, which contributes to reshaping today's power system toward a sustainable and clean energy system. However, due to the uncertainty of renewable energy generation and load, this has caused a strong obstacle to day-ahead dispatching plans of a microgrid\cite{kumar2020optimal}. It's known that an accurate forecasting result can provide a reliable basis for dispatching plans to arrange the start and stop of microturbines {\color{black}(MT)} and set the spinning reserve capacity. A scheduling model combined with advanced forecasting methods can undoubtedly reduce the impact of the uncertainty of renewable generations and load on microgrids, and improve the economy of microgrid operation\cite{aguera2018weather}.
\subsection{Literature Review}
Forecasting methods of renewable energy generation and load have been extensively investigated. As far as current forecasting methods are concerned, they are mainly divided into two categories. 1) One is the traditional time series analysis method: In the 1970s, George Box and Gwilym Jenkins proposed the ``Box-Jenkins method'', Ref.\cite{vahakyla1980short} used this method for the first time for short-term load forecasting, and subsequently, more modern mathematical theories were applied to power system forecasting. In  \cite{lee2011forecasting}, a gray theory was used to predict electricity prices and renewable energy power generation, and \cite{wang2018conditional} used probability forecasting methods to predict the load. However, traditional time series forecasting methods heavily rely on the choice of model parameters, the appropriate model parameters can largely determine the accuracy of the predicted results, and they have a poor generic capability. 2) Another one is machine learning (ML) algorithms such as support vector machines and artificial neural networks: Ref.\cite{jiang2016short} used support vector regression to finely predict the load of the distribution network and an artificial neural network (ANN) was used to forecast photovoltaic (PV) power output in \cite{rodriguez2018predicting}; With the development of computer technology and the advent of the era of big data, deep learning has become a hot spot in forecasting methods; In \cite{tan2019ultra}, long short term memory (LSTM) network was used to predict load, and a more sophisticated neural network was proposed in \cite{afrasiabi2019multi}, which combined convolutional neural network (CNN) with gated recurrent unit (GRU) to predict renewable energy outputs, electricity price, and load. {\color{black}But traditional supervised learning needs high-quality datasets for training to obtain a perfect forecasting performance, which limits the application of such methods in real-world systems to a certain extent.} Reinforcement learning, which has received widespread attention, has also been applied to building energy consumption forecasting \cite{liu2020study}. {\color{black} Compared with traditional time series analysis methods, forecasting methods based on machine learning can provide more accurate prediction results, but such kind of methods require a lot of domain knowledge and human interventions when constructing a forecasting model, which seriously affects the efficiency of model construction and the model's versatility.} As is known, accurate multi-period forecasting results are necessary for microgrid day-ahead scheduling. Unfortunately, these methods used in the aforementioned literature can only guarantee a certain accuracy for multi-period prediction issues, which poses a significant barrier  for developing a forecasting model with high performances in practical applications. In this context, our motivation is to leverage reinforcement learning framework for renewable energy and load forecasting so that compared with supervised learning, {\color{black} which is more robust and} more potential connections can be learned under the same amount of information to improve the performance of a forecasting model\cite{kuremoto2019training}.
\par For isolated microgrids, how to deal with the uncertainty of renewable energy output and load is a crucial problem due to its relatively small capacity and unavailability to obtain the power support from the main grids. An economic operation model of microgrid was proposed in \cite{hernandez2005fuel} earlier, on the basis of including multiple forms of DGs, the model considered the constraints of microgrid cogeneration, reserve capacity, etc. Ref.\cite{ebrahimi2019adaptive} adopted the method of robust programming, expressed the variation range of wind turbine (WT) output and load according to the uncertainty set. In order to improve the operating economy of the microgrid, Ref.\cite{wang2020quantitative} used multivariate global sensitivity analysis to identify and retain the key uncertain factors that affect the operation of the system. In \cite{ciftci2019data}, authors fitted the best probability distribution based on historical data of WT output and load, then used the chance constraint programming method for microgrid dispatch. This method balanced the economy and reliability of microgrid operation, but it used  WT output and load data in their expectation form as the input for scheduling, which {\color{black}was not able} to provide a reliable basis for the dispatch model. In \cite{wen2019optimal}, authors used LSTM to predict PV output and load, and then used particle swarm optimization to optimize the scheduling model. However, in this method, the forecasting model and the scheduling model are two separate parts.
\par It can be seen from the above literature that the existing research has made an in-depth discussion on the optimal scheduling of microgrids, however, there are still the following gaps that need to be addressed:
\begin{enumerate}
    {\color{black} \item The architecture and hyperparameters of a machine learning model have a significant impact on model performances, and building a learning model requires lots of domain knowledge and human interventions.} 
    \item The prediction accuracies of most  existing forecasting models are not ideal for multi-period prediction, which cannot provide reliable data support for microgrid day-ahead scheduling.
    \item The existing forecasting approaches rarely consider uncertainty modeling of forecasting errors and the impact of the errors on forecasting results, which will inevitably deteriorate the forecasting performance.
    \item Most of current microgrid scheduling models adopt the load and renewable energy generations in the form of fixed values or rough estimations, and do not combine a sophisticated forecasting model to establish an organic dynamic integration.
  
\end{enumerate}

\subsection{Contribution of This Paper} 
The main contributions of this paper are the following fourfold:
\begin{enumerate}
    {\color{black}\item  A prioritized experience replay automated reinforcement learning (PER-AutoRL) is designed to predict renewable energy outputs and load in a customized manner, which can automatically determine the most appropriate model architecture and optimize hyperparameters based on the input data. This design will improve modeling efficiency, strengthen the combination of forecasting and scheduling, and provide reliable data support for scheduling.
    \item We propose a multi-period single-step forecasting method based on PER-AutoRL, which can significantly improve the forecasting accuracy by solving the error accumulation issues suffered by traditional multi-step forecasting methods.
    \item By modeling uncertainty of forecasting errors, we consider the impact of the errors on the prediction accuracy, and revise the predicted values according to the error distribution described by the t location-scale (TLS) distribution to further improve the forecasting performance.
    \item We construct a microgrid scheduling model integrating the designed forecasting method with consideration of demand response, which can significantly reduce the microgrid operating costs.}
\end{enumerate}

\section{Forecasting Model based on {\color{black}Prioritized Experience Replay Automated} REINFORCEMENT LEARNING}\label{sec.DDPG MODEL}

\subsection{Deep Deterministic Policy Gradient}
Deep reinforcement learning {\color{black}(DRL)} is a combination of deep learning and reinforcement learning, {\color{black}which provides solutions for the perception and decision-making problems of sophisticated systems \cite{mnih2015human}.} As a powerful DRL algorithm, deep deterministic policy gradient (DDPG) has demonstrated a powerful ability in dealing with continuous action space problems. 
\par In DDPG, deep neural networks with parameters $\theta^\mu$ and $\theta^Q$ are used to represent the main actor network, $a=\pi(s|\theta^\mu)$ and the main critic network, $Q(s,a|\theta^Q)$. The main function of the Actor part is to interact with the environment, that is, directly select the output action, $a$ according to state, $s$, 
and get the next {\color{black}state, $s'$} and reward, $r$ after interacting with the environment. The objective function is the total reward with a discount factor $\gamma$, which is formulated as
\begin{equation}\label{eq.object function}
    \begin{aligned}
       J({\theta ^\mu }) = {E_{{\theta ^\mu }}}[{r_1} + \gamma {r_2} + {\gamma ^2}{r_3} + ... + {\gamma ^{i - 1}}{r_i}]
    \end{aligned}
\end{equation}
\par To improve the total reward $J$, the objective function is optimized through the stochastic gradient. Silver et al. proved that the gradient of the objective function with respect to $\theta^\mu$ is equivalent to the expectation gradient of the Q-value function with respect to $\theta^Q$ \cite{silver2014deterministic}, when updating the main actor network, the gradient can be approximated as

\begin{equation}\label{eq.update main actor network}
    \begin{aligned}
       \nabla _{{\theta ^\mu }}J \approx \frac{1}{N}\sum\limits_i {[\nabla {}_aQ(s,a|{\theta ^Q}){|_{s = {s_i},a = \pi ({s_i})}}\nabla {}_{{\theta ^\mu }}\pi (s|{\theta ^\mu })|s = {s_i}]}
    \end{aligned}
\end{equation}

\par The main function of the Critic part is to evaluate the strategy proposed by the Actor. The main critic network is updated with the goal of minimizing the loss function, the loss function is as follows:
{\color{black}\begin{equation}
    L({\theta ^Q}) = {E_{s,a,r,s'D}}(TD-error)^2
\end{equation}
\begin{equation}
   TD-error = [r + \gamma Q'(s',\pi (s'|{\theta ^{\mu '}})|{\theta ^{Q'}})] - Q(s,a|{\theta ^Q}) 
\end{equation}}{\color{black}where $Q'$ is the target Q-value function; $\theta ^{\mu '}$ and $\theta ^{Q'}$ denote the parameters of the target actor network and the target critic network, respectively.}

\par The target networks of the Actor and Critic part are constructed mainly to make the network training more stable, their parameters can be updated with a soft update factor $\varsigma$ every fixed period of time:
\begin{equation}\label{eq.update target networks}
    \begin{aligned}
     \begin{array}{l}
{\theta ^{Q'}} \leftarrow \varsigma {\theta ^Q} + (1 - \varsigma ){\theta ^{Q'}}\\
{\theta ^{\mu '}} \leftarrow \varsigma {\theta ^\mu } + (1 - \varsigma ){\theta ^{\mu '}}
\end{array}
    \end{aligned}
\end{equation}

{\color{black}\subsection{PER-AutoRL}
Automated machine learning (AutoML) can be regarded as a system with powerful learning and generalization capabilities on given data and tasks. For traditional machine learning, obtaining a model with good performances often requires a lot of experts' experience and human debugging. AutoML is designed to reduce the demand for data scientists and enable domain experts to automatically build ML applications without much requirement for statistical and ML knowledge, it can undoubtedly improve the efficiency of model design and reduce the difficulty in applying ML.

\par To simplify the deployment of DRL models and improve the efficiency, we propose a PER-AutoRL by extending AutoML to DRL. We use the Metis method, a type of sequence-based Bayesian optimization algorithm, to determine the architecture of the DRL model and optimize the hyperparameters \cite{li2018metis}. By using the DDPG algorithm as an overall framework, the Actor part makes predictions based on the input data and the Critic part evaluates the prediction results and obtains the optimal prediction policy in this study, which enables the perception ability of deep learning to integrate the decision-making ability of reinforcement learning for solving the forecasting problem. We set the actor network and the critic network to a fix depth feed-forward fully-connected neural network, and then determine the network architecture according to the size of each layer and the activation function, and then optimize the hyperparameters in the model. 
\par For DRL, an appropriate reward function can improve the performance of the model. In this respect, we set a reward function pool to store a reward function that may be applicable and treat the reward function as a hyperparameter, which is optimized in synchronization with other hyperparameters. For the prediction problem studied in this paper, the absolute error $\sigma$, mean absolute error (MAE), mean square error (MSE), mean absolute percentage error (MAPE), root mean square error (RMSE) and the coefficient of determination $R^2$ are slightly modified as options of the reward function. Their expressions are
\begin{equation}
 \left\{ \begin{array}{l}
 - |\sigma | =  - |y_i^{ac} - {p_i}|\\
 - MAE =  - {{[\sum\nolimits_{i = 1}^N {(|y_i^{ac} - {p_i}|)]} } \mathord{\left/
 {\vphantom {{[\sum\nolimits_{i = 1}^N {(|y_i^{ac} - {p_i}|)]} } N}} \right.
 \kern-\nulldelimiterspace} N}\\
 - MSE =  - {{[\sum\nolimits_{i = 1}^N {{{(y_i^{ac} - {p_i})}^2}]} } \mathord{\left/
 {\vphantom {{[\sum\nolimits_{i = 1}^N {{{(y_i^{ac} - {p_i})}^2}]} } N}} \right.
 \kern-\nulldelimiterspace} N}\\
 - MAPE =  - {{[\sum\nolimits_{i = 1}^N {|(y_i^{ac} - {p_i})/y_i^{ac}|]} } \mathord{\left/
 {\vphantom {{[\sum\nolimits_{i = 1}^N {|(y_i^{ac} - {p_i})/y_i^{ac}|]} } N}} \right.
 \kern-\nulldelimiterspace} N}\\
 - RMSE =  - \sqrt {{{[\sum\nolimits_{i = 1}^N {{{(y_i^{ac} - {p_i})}^2}]} } \mathord{\left/
 {\vphantom {{[\sum\nolimits_{i = 1}^N {{{(y_i^{ac} - {p_i})}^2}]} } N}} \right.
 \kern-\nulldelimiterspace} N}} \\
{R^2} = 1 - \left\{ {{{[\sum\nolimits_{i = 1}^N {{{(y_i^{ac} - {p_i})}^2}]} } \mathord{\left/
 {\vphantom {{[\sum\nolimits_{i = 1}^N {{{(y_i^{ac} - {p_i})}^2}]} } {[\sum\nolimits_{i = 1}^N {{{(y_i^{ac} - \overline {y_i^{ac}} )}^2}]} }}} \right.
 \kern-\nulldelimiterspace} {[\sum\nolimits_{i = 1}^N {{{(y_i^{ac} - \overline {y_i^{ac}} )}^2}]} }}} \right\}
\end{array} \right.
\end{equation}
where $N$ is the number of samples, $y^{ac}$ is the actual value, $\overline {y^{ac}}$ is the mean value of $y^{ac}$, and $p$ is the prediction result.

\begin{algorithm}[!t]\label{code} 
\small
\color{black}
\renewcommand{\algorithmicrequire}{\textbf{Initialize:}}
\renewcommand{\algorithmicensure}{\textbf{return}}
    \caption{AutoRL with Prioritized Experience Replay} 
    \label{alg:code}
	\begin{algorithmic}[1] 
    \REQUIRE Neural network architecture $\theta^\mu$, $\theta^Q$.
    \REQUIRE Replay buffer $R$ with size $S$, reward function.
    \REQUIRE Maximum priority, parameters $\iota$, $\beta$.
    \FOR{trail = 1, ..., $M$}
    \STATE {\color{black}Select new neural network architecture $\theta^\mu$, $\theta^Q$ according to the Metis Tuner.}
    \STATE Select reward function from reward function pool${}$.
    \STATE Select an initial state $s_t$ from state space.
    \FOR{$t$=1, ..., $H$} 
    \STATE Select action $a_t$ according to the new policy.
    \STATE Obtain reward $r_t$ and new state $s_{t+1}$.
    \STATE  Store experience $(s_t, a_t, r_t, s_{t+1})$ in replay buffer $R$ and set $D_t = \max_{i<t} D_i$.
    \IF{$t>S$}
    \FOR{$j$=1, ..., $N$}
    \STATE Sample experience $j$ with probability $P(j)$.
    \STATE Compute importance-sampling weight $W_j$ and TD-error $\delta_j$.
    \STATE Update the priority of transition $j$ according to absolute TD-error $|\delta_j|$.
    \ENDFOR
    \STATE Update main critic network according to minimize the loss function: $L = \frac{1}{N}\sum\nolimits_i {{w_i}\delta _i^2}$. 
    \STATE Update main actor network according to (\ref{eq.update main actor network}).
    \STATE Update target networks according to (\ref{eq.update target networks}).
    \ENDIF
    \ENDFOR
    \STATE {\color{black}Collect the testing MAPE and upload it to the Metis Tuner}
    \ENDFOR
    \STATE Select the best neural network architecture $\theta^\mu$, $\theta^Q$ and the best policy $\pi$ according to minimal MAPE.
    \STATE \textbf{return} $\theta^\mu$, $\theta^Q$,$\pi$
  \end{algorithmic}
\end{algorithm}

\par Compared with ML, DRL has more hyperparameters that need to be optimized, which will affect the efficiency of the AutoRL. This work adopts Prioritized Experience Replay (PER) to improve the experience replay mechanism in DDPG, which greatly reduces the  training time \cite{hou2017novel}. The core of PER is to increase the frequency of valuable experiences. A repeated replay of extreme experience helps accelerate the agent recognize how to choose the right action to obtain high reward, or avoid the terrible results of choosing the wrong action. The DDPG updates the critic network according to TD-errors. the larger its absolute value, the more aggressive the correction of the critic network, which means that the experience has a higher value. Therefore, this paper selects the absolute value of TD-errors as the experience evaluation index to rank the experience and defines the probability $P(j)$ of the sampled experience $j$ according to this rank.
\begin{equation}
    P(j) = \frac{{D_j^\iota }}{{\sum\nolimits_{j=1}^N {D_j^\iota } }}
\end{equation}
where $D_j=\frac{1}{rank(j)}$, $rank(j)$ is the rank of experience $j$, $N$ is the total number of experience stored in the reply buffer, $\iota$ is a parameter controlling the  prioritization. 
\par To avoid frequent sampling of high TD-errors experiences that may cause the neural network to oscillate or diverge, we set the importance-sampling weight for the critic network updates:
\begin{equation}
    {W_j} = \frac{1}{{{S^\beta } \cdot P{{(j)}^\beta }}}
\end{equation}
where $S$ is the size of the reply buffer, parameter $\beta$ controls to what extent the correction is used. 
\par The integrated algorithm of PER-AutoRL is shown in Algorithm \ref{alg:code}.
}

\subsection{Multi-period Forecasting Based on {\color{black} PER-AutoRL}}\label{sec.FM}
The multi-period forecasting methods can be divided into two categories: multi-step forecasting and single-step forecasting. The former takes the prediction results of each step to make next prediction; {\color{black}while the latter is based on the real value in the previous period to predict the next-period value; after each prediction step, the real value in the previous period needs to be updated for the next-period prediction.} The single-step forecasting cannot provide complete prediction results for day-ahead scheduling, because the real values of each period cannot be updated timely.
And it should be noted that the errors accumulation is inevitable in multi-step forecasting, resulting in a decrease in the forecasting accuracy.
\par This paper proposes a multi-period forecasting method based on PER-AutoRL, which transforms the multi-step forecasting problem into a multi-period single-step forecasting problem. The proposed method addresses the error accumulation issue and improves the prediction accuracy, the specific procedure is as follows:
\begin{enumerate}
    {\color{black}
    \item Dataset preprocessing. Extract the data in the same period of a day from the dataset and reconstruct it into multiple sets of new time series (this work uses 1h as the time interval to reconstruct the dataset into 24 new time series).}
    {\color{black}\item Constructing PER-AutoRL model. {\color{black}The different architecture and hyperparameters of the PER-AutoRL model for predicting each time series are automatically determined according to the separate new time series.} For the prediction problem in this study, we select the data of the same period in 7 consecutive days to construct a matrix as the state $s$, and then the agent outputs a predicted value as the action $a$ based on its observed state. The loss function of the main actor network aims to minimize the forecasting errors, while the loss function of the main critic network is built for minimizing the gaps between the predicted value and the true value of the reward.}
    \item Training PER-AutoRL model. The weight matrices and biases of the main actor network and the main critic network are determined until the predetermined number of training episodes is completed.
    \item Day-ahead forecasting. Respectively forecast the values in each period, then integrate the forecasting results of each group  to form a forecasting data for the next 24 hours.
    {\color{black}\item Revising forecasting results. After obtaining the error distribution via the forecasting errors of PER-AutoRL, we calculate the expected values of errors, then revise the forecasting values based on them. The revised forecasting results are used as the input of the scheduling model and the error expected values of PER-AutoRL are prepared for the determination of the spinning reserve in the scheduling.}
\end{enumerate}

{\color{black}By constructing a multi-period single-step prediction model based on PER-AutoRL, the error accumulation is effectively solved. In addition, benefits from the training methods and exploration capabilities of DRL, the robustness and practicability of the prediction model have been improved\cite{kuremoto2019training}}.

\subsection{Uncertainty Modeling of Forecasting Errors and Revised Forecasting Results}
\subsubsection{Stochastic Model of Forecasting Errors}
In this paper, TLS distribution is used to describe the probability distribution of the PER-AutoRL prediction errors \cite{wang2019novel}:
\begin{equation}
    \begin{aligned}
    f(x;\mu ,\varepsilon ,\vartheta ) = \frac{{\Gamma (\frac{{\vartheta  + 1}}{2})}}{{\varepsilon \sqrt {\vartheta \pi } \Gamma (\frac{\vartheta }{2})}}{[\frac{{\vartheta  + {{(\frac{{x - \mu }}{\varepsilon })}^2}}}{\vartheta }]^{ - \vartheta  + 1/2}}  
    \end{aligned}
\end{equation}
\begin{figure}[t]
	\centering	
	\includegraphics[width=2.6in]{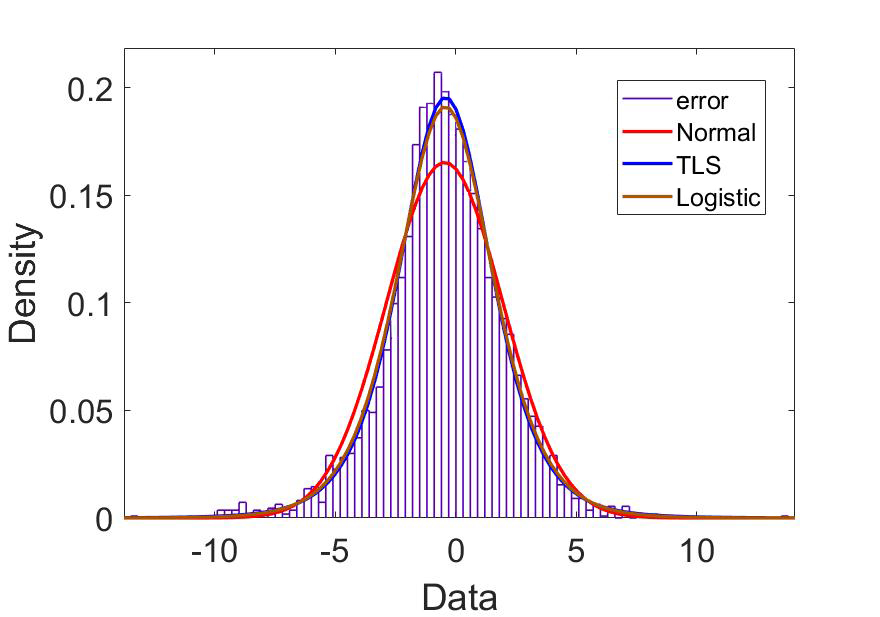}	
	\caption{Probability distribution of forecasting errors.}	\label{fig_TLS}
\end{figure}
where $\Gamma$ is the gamma function, $\mu$ is the mean value, $\varepsilon$ is the standard deviation, and $\vartheta$ is the shape coefficient.

Fig. \ref{fig_TLS} is a fitting diagram of the probability distribution of WT output forecasting errors for a total of one year from January 1, 2018 to December 31, 2019. It can be seen from Fig. \ref{fig_TLS} that compared with the normal distribution and logistic distribution, the TLS distribution can better describe the forecasting error probability distribution.

\subsubsection{Revised Forecasting Results}
To obtain a more accurate forecasting result, we acquire the expectation of error via the distribution of the prediction errors, and then modify the forecasting results. The calculation process is:
\begin{equation}\label{eq.RFR}
    \begin{aligned}
\left\{  
             \begin{array}{lr}  
             C(P_{L,t}^{Pred}) = P_{L,t}^{Pred} - E(\sigma _t^L)  \vspace{1ex} \\
             C(P_{WT,t}^{Pred}) = P_{WT,t}^{Pred} - E(\sigma _t^{WT}) \vspace{1ex} \\ 
             C(P_{PV,t}^{Pred}) = P_{PV,t}^{Pred} - E(\sigma _t^{PV}) 
             \end{array}  
\right. 
      \end{aligned}
\end{equation}
where $C(P_{L,t}^{Pred}),C(P_{WT,t}^{Pred}),C(P_{PV,t}^{Pred})$ are the revised forecasting results of load, WT and PV outputs, respectively;  $P_{L,t}^{Pred},P_{WT,t}^{Pred},P_{PV,t}^{Pred}$ are the forecasting results of load, WT and PV outputs; {\color{black}$\sigma _t^L$, $\sigma _t^{WT}$, $\sigma _t^{PV}$ are the forecasting errors of load, WT and PV outputs;} and $E(\sigma _t^L),E(\sigma _t^{WT}),E(\sigma _t^{PV})$ are the expectation of forecasting errors of load, WT and PV outputs, respectively.

\subsubsection{Evaluation Indices}
This paper evaluates the forecasting methods based on the common prediction evaluation indicators MAPE and RMSE\cite{li2019short}.

\section{Serialization Modelling of Random Variables}\label{sec.SOT}
\subsection{Sequence Operation Theory}
Sequence operation theory {\color{black}(SOT)} is a powerful mathematical tool for dealing with multiple uncertain variables. The core idea is to acquire a probabilistic sequence of random variables through discretization, and then generate a new sequence through sequence operations\cite{kang2002sequence}.
\par Suppose the discrete sequence $a(i)$ of length $N_a$ satisfies the following conditions:
\begin{equation}
    \begin{aligned}
      \sum\limits_{{i_a} = 0}^{{N_a}} {a({i_a})}  = 1,{\rm{  }}a({i_a}) \ge 0,{\rm{  }}i = 0,1,2,...,N{}_a
    \end{aligned}
\end{equation}
the sequence is regarded as  a probabilistic sequence.

\subsection{Sequence Description of Equivalent Load Forecasting Errors}
During period $t$, the forecasting errors $\sigma^{WT}$, $\sigma^{PV}$ and $\sigma^{L}$ of WT, PV outputs, and load are all random variables. In this study, it is assumed that the uncertainty of them does not affect each other. {\color{black}After discretizating the continuous probability distributions of WT outputs forecasting errors according to the step length $q$, the probabilistic sequences $a(i_{a, t})$ can be obtained by 
\begin{equation}
    \begin{aligned}
    \begin{array}{l}
  
a({i_{a,t}}) = \left\{ \begin{array}{l}
  \vspace{0.1cm}
\int_{\sigma _{\min }^{WT,t}}^{\sigma _{\min }^{WT,t} + q/2} {{f_o}({\sigma ^{WT}})d} {\sigma ^{WT}},\quad {i_{a,t}} = 0\\ \vspace{0.1cm}
\int_{\sigma _{\min }^{WT,t} + {i_{a,t}}q - q/2}^{\sigma _{\min }^{WT,t} + {i_{a,t}}q + q/2} {{f_o}({\sigma ^{WT}})d} {\sigma ^{WT}},\\ \qquad \qquad \qquad \qquad \qquad \quad{i_{a,t}}> 0, {i_{a,t}} \ne {N_{a,t}}\\ \vspace{0.4cm}
\int_{\sigma _{\min }^{WT,t} + {i_{a,t}}q - q/2}^{\sigma _{\min }^{WT,t} + {i_{a,t}}q} {{f_o}({\sigma ^{WT}})d} {\sigma ^{WT}},\ {i_{a,t}} = {N_{a,t}}
\end{array} \right.\\
\qquad \qquad \qquad \qquad \  \quad{N_{a,t}} =[ \frac{{\sigma _{\max }^{WT,t} - \sigma _{\min }^{WT,t}}}{q}]
    \end{array}
    \end{aligned}
\end{equation}
where $f_o$ is the probability density function, $N_{a,t}$ is the length of the sequence, $\sigma _{\min }^{WT,t}$ and $\sigma _{\max }^{WT,t}$ are the minimum and maximum values of WT outputs forecasting errors.
\par In the same way, the probabilistic sequence $b(i_{b,t})$, $d(i_{d, t})$ with length $N_{b,t}$, $N_{d,t}$ of the PV outputs forecasting errors and load forecasting errors can be obtained.}
\par We define the equivalent load (EL) as the difference between the load power and the joint outputs of WT and PV. {\color{black}The probabilistic sequence $c(i_c, t)$ of the joint outputs forecasting errors with length $N_{c,t}$ }can be obtained by addition-type-convolution operation:
{\color{black}\begin{equation}
    \begin{aligned}
    \begin{array}{l}
c({i_{c,t}}) = a(i_{a,t}^{}) \oplus b({i_{b,t}}) = \sum\nolimits_{{i_{a,t}} + {i_{b,t}} = {i_{c,t}}} {a(i_{a,t}^{})b({i_{b,t}})} ,\\
\ \ \ \ \ \ \ \ \ \ \ \ \ \ \ {i_{c,t}} = 0,1,...,{N_{a,t}} + {N_{b,t}}
    \end{array}
    \end{aligned}
\end{equation}}{\color{black}Then, subtraction-type-convolution is employed to calculate the probabilistic sequence $e(i_e, t)$ of the EL forecasting errors\cite{li2018optimal}:}
{\color{black}
\begin{equation}
\begin{array}{l}
e({i_{e,t}}) = d({i_{d,t}}) \ominus c({i_{c,t}})\\
\qquad\ \  =\left\{ \begin{array}{l}
\sum\nolimits_{{i_{d,t}} - {i_{c,t}} = {i_{e,t}}} {d({i_d})} c({i_c}),1 \le {i_{e,t}} \le {N_{e,t}}\\
\sum\nolimits_{{i_{d,t}} \le {i_{c,t}}} {d({i_d})} c({i_c}),{i_{e,t}} = 0
\end{array} \right.
\end{array}
\end{equation}

The corresponding relationship between prediction errors of EL and their probabilistic sequence are shown in Tab. \ref{tab.PS2}, where $\sigma _{\min }^{EL,t}$ denotes the minimum value of EL forecasting errors, $N_{e,t}$ is the length of $e({i_{e,t}})$. The expectation of $e({i_{e,t}})$ can be calculated by 
\begin{equation}
    \begin{aligned}
     E(e) = \sum\limits_{i_{e,t} = 0}^{{N_{e,t}}} {[(\sigma _{\min }^{EL,t}+i_{e,t}q)\cdot e(i_{e,t})]}
    \end{aligned}
\end{equation}}

\begin{table}[t]
\caption{ {\color{black}Prediction error of the EL and its probabilistic sequence}}\label{tab.PS2}
\centering
\renewcommand\arraystretch{1.8}
\resizebox{1\columnwidth}{!}{
\begin{tabular}{rclllll}
\hline
Error(kW)    &{\color{black}$\sigma _{\min }^{EL,t}$}    &{\color{black}$\sigma _{\min }^{EL,t}+q$}    &{\color{black}...}  &{\color{black}$\sigma _{\min }^{EL,t}+i_{e,t}q$}   &{\color{black}...} &{\color{black}$\sigma _{\min }^{EL,t}+N_{e,t}q$}\\
\hline
Probability  &$e(0$) &\quad\ $e(1)$ &...  &\quad\ $e(i_{e,t})$ &... &\quad\ $e(N_{e,t})$\\
\hline
\end{tabular}}
\end{table}

\section{Optimal Scheduling Model of Microgrid Based on {\color{black}PER-AutoRL} Forecasting}\label{sec.scheduling model}
\subsection{Optimal Scheduling Model}\label{sec.model}
\subsubsection{Objective Function}
The objective function $F_c$ of the optimal scheduling model considering demand response is constructed to minimize the total microgrid operating costs that are comprised of the MT units fuel costs and the spinning reserve costs. {\color{black}In light of the nature of the demand-side loads, this work divides electric load into fixed load and interruptible load. Since the load interruption inevitably affects user experience, certain subsidies will generally be provided to users \cite{li2021optimal}, and the subsidies are also considered in the total operating costs. The objective function is as follows:}
{\color{black}\begin{equation}
   \begin{array}{l}
\min {F_c} = \sum\limits_{t = 1}^T {{\rm{[}}\sum\limits_{n = 1}^{{M_G}} {({\varsigma _n}R_{n,t}^{MT} + {\tau _n}{S_{n,t}} + \kappa P_t^{IE}} } \\
\qquad \qquad \qquad \qquad \qquad \quad + {U_{n,t}}({\psi _n} + {\xi _n}P_{n,t}^{MT}))]
    \end{array}
\end{equation}}where $T$ denotes the total number of time period $t$ in a scheduling cycle ($T$=24 in this paper); $M_G$ denotes the total number of MT units; $n$ is the MT number; $\psi_n$ and $\xi_n$ represent the consumption factors of MT $n$; $U_{n, t}$ and $S_{n, t}$ are 0-1 variables representing the state variable and the startup variable of MT $n$, respectively; {\color{black}$P_t^{IE}$ is the interruptible electric load, and $\kappa$ is the subsidies;} $\varsigma_n$ and $\tau_n$ denote the costs of spinning reserves and the startup costs of MT $n$; $P_{n,t}^{MT}$ and $R_{n,t}^{MT}$ are respectively the output power and the spinning reserve of MT $n$ in period $t$.
\subsubsection{Constraint Conditions}
\paragraph{Power balance constraint} To maintain the power balance of the isolated microgrid, the controllable loads need to be deployed, so the power balance constraints are
{\color{black}\begin{equation}\label{eq.PB}
    \begin{aligned}
\sum\limits_{n = 1}^{{M_G}} {P_{n,t}^{MT} + } P_t^{DC} - P_t^{CH} = C(P_{EL,t}^{Pred}) \\
\qquad \qquad \qquad \qquad \qquad + P_{n,t}^{CNLOAD}-P_t^{IE} 
    \end{aligned}
    \qquad \forall t
\end{equation}
\begin{equation}
P_t^{IE} \le \rho C(P_{EL,t}^{Pred}) \quad \forall t
\end{equation}}
\begin{equation}
    \begin{aligned}
P_{EL,t}^{Pred} = P_{L,t}^{Pred} - (P_{WT,t}^{Pred} + P_{PV,t}^{Pred})
    \end{aligned}
\end{equation}
where {\color{black}$\rho$ is the ratio of $P_t^{IE}$,} $P_t^{CH}$ and $P_t^{DC}$ are the charging and discharging power of the energy storage in period $t$, $P_t^{CNLOAD}$ is the controllable load. $P_{EL,t}^{Pred}$ is the predicted value of the EL, $C(P_{EL,t}^{Pred})$ is the revised value of $P_{EL,t}^{Pred}$, the calculation process is as follows:
{\color{black}\begin{equation}\label{eq.cel}
    \begin{aligned}
\begin{array}{l}
C(P_{EL,t}^{Pred}) = P_{L,t}^{Pred}-E(\sigma _t^L)-[P_{WT,t}^{Pred}-E(\sigma _t^{WT})] \\
\qquad \qquad \quad \  -[P_{PV,t}^{Pred} - E(\sigma _t^{PV})]
\end{array}
    \end{aligned}
\end{equation}}

\paragraph{MT output constraint}
The output of MT must comply with the following inequality:
\begin{equation}
    \begin{aligned}
{U_{n,t}}P_{n,\min }^{MT} \le P_{n,t}^{MT} \le {U_{n,t}}P_{n,\max }^{MT}\ {\rm{  }}\forall t,n \in {M_G}
    \end{aligned}
\end{equation}
where $P_{n,\max }^{MT}$ and $P_{n,\min }^{MT}$ are the upper and lower limits of the output power of MT unit $n$.

\paragraph{Energy storage system constraints}
In this paper, the energy storage system (ESS) adopts lead-acid batteries to balance the random fluctuations in the microgrid because it provides many advantages, such as low price and long service life \cite{neto2018dual}.
\par Charge-discharge equation: The relationship between the ESS and the charge-discharge powers is expressed as
\begin{equation}
    \begin{aligned}
{S_{t + 1}} = {S_t} + ({\eta _{ch}}P_t^{CH} - P_t^{DC}/{\eta _{dc}})\Delta t\quad{\rm{   }}\forall t
    \end{aligned}
\end{equation}
where $S_{t+1}$ and $S_t$ are respectively the ESS energy storage at the beginning of period $t+1$ and $t$, $\eta_{ch}$ and $\eta_{dc}$ are the charge/discharge efficiency, and $\Delta t$ denotes the duration of each period, which is taken as 1h in this paper.
\par The output limits of lead-acid batteries:
\begin{equation}
    \begin{aligned}
\left\{  
             \begin{array}{lr}  
             0 \le P_t^{DC} \le P_{\max }^{DC}\quad{\rm{  }}\forall t \\
             0 \le P_t^{CH} \le P_{\max }^{CH}\quad{\rm{  }}\forall t  
             \end{array}  
\right. 
      \end{aligned}
\end{equation}
where $P_{\max }^{CH}$ and $P_{\max }^{DC}$ are the maximum values of lead-acid battery charge and discharge in time period $t$, respectively.
\par The capacity limit of lead-acid batteries is
\begin{equation}
    \begin{aligned}
{S_{\min }} \le {S_t} \le {S_{\max }}\quad{\rm{   }}\forall t
    \end{aligned}
\end{equation}
where $S_{max}$ and $S_{min}$ are the maximum and minimum allowable capacities of the batteries, respectively.
\par The ESS starting and ending constraints are
\begin{equation}
    \begin{aligned}
{S_0} = {S_{{T_{end}}}} = {S_*}\quad\forall t
    \end{aligned}
\end{equation}
where $S_0$ is the ESS initial energy storage, $S_*$ is the limit of the initial energy storage in ESS, and $T_{end}$ is the end of the scheduling cycle ($T_{end}$=24h in this paper). For the ESS energy balance, we set the remaining capacity $S_{T_{end}}$ of the ESS after each scheduling cycle the same as $S_0$.
\paragraph{Spinning reserve constraint}
The spinning reserve is an important resource for an isolated microgrid to suppress the renewable DGs outputs fluctuations and ensure the system operates reliably. The required spinning reserves are provided by the MT units and ESS. Therefore, the spinning reserve constraints are expressed as 
\begin{equation}
    \begin{aligned}
P_{n.t}^{MT} + R_{n,t}^{MT} \le {U_{n,t}}P_{n,\max }^{MT}\quad\forall t,n \in {M_G}
    \end{aligned}
\end{equation}
\begin{equation}
    \begin{aligned}
P_{{\mathop{\rm Re}\nolimits} ss,{\rm{ }}t}^{} \le \min \{ {\eta _{dc}}({S_t} - {S_{\min }})/\Delta t,P_{\max }^{DC} - P_t^{DC}\} \quad\forall t
    \end{aligned}
\end{equation}
where $P_{Ress, t}$ denotes the ESS reserve capacity.
\par {\color{black}It can be seen from (\ref{eq.PB}) that when the power balance constraint is considered, the EL is processed by the revised prediction value $C(P_{EL,t}^{Pred})$. Due to the uncertainty of load and renewable energy outputs, the total spinning reserve provided by ESS and MTs must be able to make up for the difference between the fluctuating EL and $C(P_{EL,t}^{Pred})$.}
\par {\color{black}When the extreme situation with zero renewable energy outputs occurs, the system needs a large spinning reserve capacity, which will incur high spinning reserve costs. However, the possibility of such cases occurring is very low.} To balance the reliability and economy in the economic operation of the microgrid, the spinning reserve constraint can be expressed as a chance constraint \cite{ciftci2019data}:
\begin{equation}\label{ep.PC}
    \begin{aligned}
{P_{rob}}\{ \sum\limits_{n = 1}^{{M_G}} {R_{n,t}^{MT} + {P_{{\mathop{\rm Re}\nolimits} ss,t}} \ge (P_{t}^L\! -\! P_{t}^{WT} \!- \!P_{t}^{PV})}  \!- C(P_{EL,t}^{Pred})\} \\
\ge \alpha {\rm{  }}\quad\forall t\qquad\qquad\qquad\qquad\qquad\quad\ 
    \end{aligned}
\end{equation}
By using
\begin{equation}
    \begin{aligned}
\left\{  
             \begin{array}{lr}  
             P_{t}^L = P_{L,t}^{Pred} - \sigma _t^L \vspace{1ex} \\
             P_{t}^{WT} = P_{WT,t}^{Pred} - \sigma _t^{WT} \vspace{1ex} \\ 
             P_{t}^{PV}= P_{PV,t}^{Pred} - \sigma _t^{PV}
             \end{array}  
\right. 
      \end{aligned}
\end{equation}
{\color{black}where $P_{t}^L$ denotes load in period $t$; $P_{t}^{WT}$ and $P_{t}^{PV}$ are the WT and PV outputs, respectively; $C(P_{EL,t}^{Pred})$ can be calculated by (\ref{eq.cel}). Accordingly, (\ref{ep.PC}) can be simplified to}
\begin{equation}\label{eq.constraint}
    \begin{aligned}
{P_{rob}}\{ \sum\limits_{n = 1}^{{M_G}} {R_{n,t}^{MT}\! +\! {P_{{\mathop{\rm Re}\nolimits} ss,t}} \ge E(\sigma _t^{EL})} \! -\! (\sigma _t^L \!- \!\sigma _t^{WT}\!-\!\sigma _t^{PV})\} \\
\ge \alpha {\rm{  }}\quad\forall t\qquad\qquad\qquad\qquad\qquad
    \end{aligned}
\end{equation}
where $\alpha$ is the preset confidence level, and $E(\sigma _t^{EL})$ is the expectation of the EL forecasting error which can be obtained by the method in Sec.\ref{sec.SOT}.

\subsection{Deterministic Transformation of {\color{black}Chance} Constraints}
To facilitate the processing of (\ref{eq.constraint}), we design a new type of 0-1 variable as follows:
{\color{black}
\begin{equation}\label{ep.0-1}
\footnotesize
    W_{{i_{e,t}}} =\left\{
\begin{aligned}
& 1,{\rm{   }}\sum\nolimits_{n = 1}^{{M_G}} {R_{n,t}^{MT} + {P_{{\mathop{\rm Re}\nolimits} ss,t}}}  \ge E(\sigma _t^{EL}) - (\sigma _{\min }^{EL,t}+i_{e,t}q) \\
& \qquad\qquad\qquad\qquad\qquad\qquad\qquad \ \ \forall t,{i_{e,t}} = 0,1,...,{N_{e,t}}\\
& 0,otherwise
\end{aligned}
\right.
\end{equation}}

\par As Tab. \ref{tab.PS2} shows, the probability of EL forecasting error {\color{black} ($\sigma _{\min }^{EL,t}+i_{e,t}q$) is $e(i_{e, t})$}, so (\ref{eq.constraint}) can be transformed into
{\color{black}
\begin{equation}\label{eq.s}
    \begin{aligned}
\sum\limits_{{i_{e,t = 0}}}^{{N_{e,t}}} {{W_{{i_{e,t}}}}e({i_{e,t}}) \ge \alpha }
    \end{aligned}
\end{equation}}

\par
\begin{figure}[t]
	\centering	
	\includegraphics[width=3.5in]{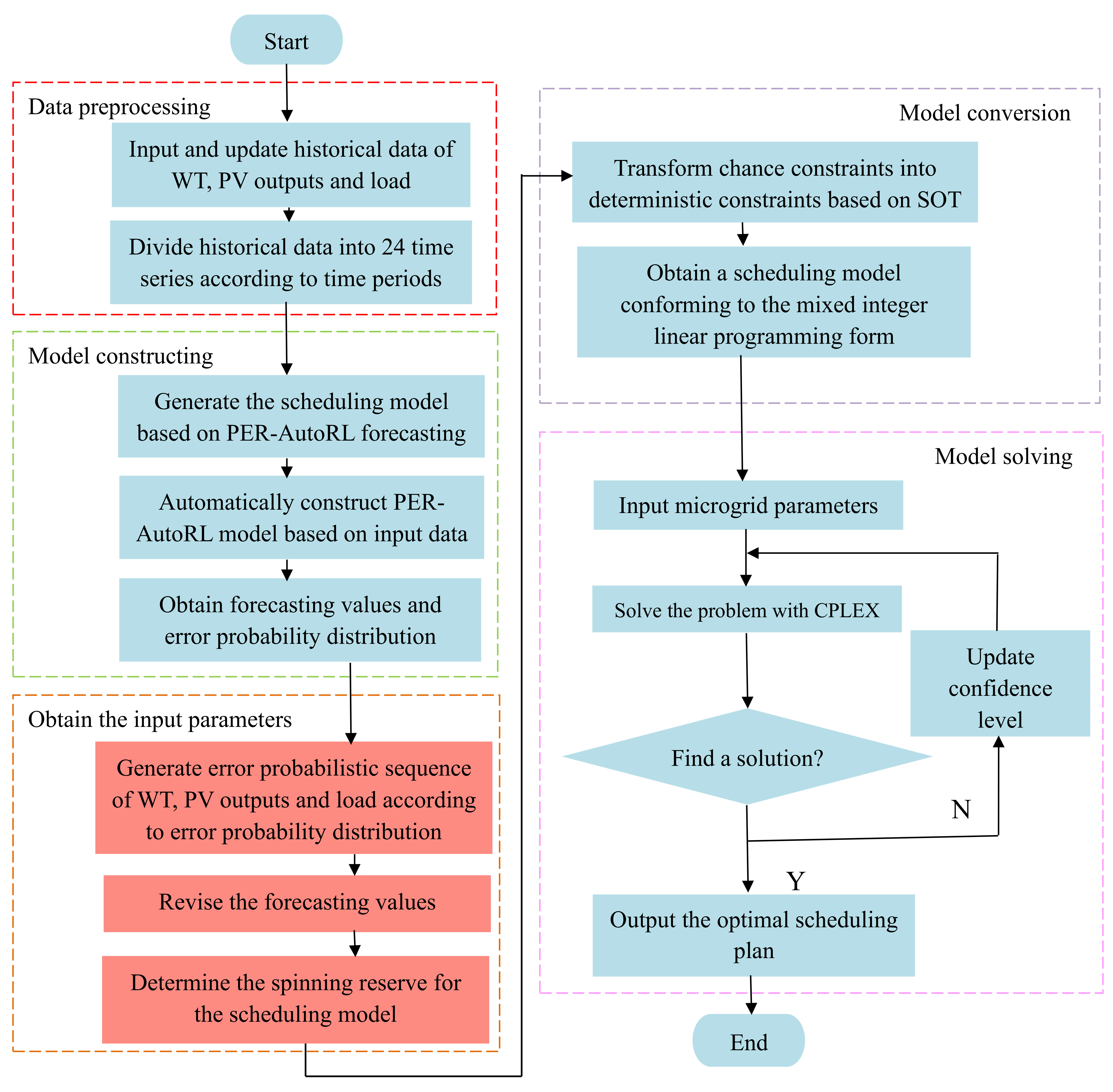}	
	\caption{{\color{black}Flowchart of the proposed scheduling model.}}	
	\label{fig.flowchart}
\end{figure}

 {\color{black}The expression of $W_{i_{e,t}}$ in (\ref{eq.s}) is not compatible with the solution form of the mixed integer programming. In order to solve the problem, (\ref{ep.0-1}) must be replaced by 
\begin{equation}\label{eq.re}
\footnotesize
    \begin{aligned}
(\sum\limits_{n = 1}^{MG} {R_{n,t}^{MT} + {P_{{\mathop{\rm Re}\nolimits} ss,t}}}  + \sigma _{\min }^{EL,t}+i_{e,t}q - E(\sigma _t^{EL}))/\Phi  \le {W_{{i_{e,t}}}} \\
\le 1 + (\sum\limits_{n = 1}^{MG} {R_{n,t}^{MT} + {P_{{\mathop{\rm Re}\nolimits} ss,t}}}  + \sigma _{\min }^{EL,t}+i_{e,t}q - E(\sigma _t^{EL}))/\Phi\\
\forall t,{i_{e,t}} = 0,1,...,{N_{e,t}}
    \end{aligned}
\end{equation}
where $\Phi$ is a very large positive number in this formula.
By substituting (\ref{eq.constraint}) with (\ref{eq.s}) and (\ref{eq.re}), the model in Sec. \ref{sec.model} will be completely transformed into the form of the mixed integer programming. The overall flowchart of the proposed scheduling model is shown in Fig. \ref{fig.flowchart} }.

\begin{figure}[t]
	\centering	
	\includegraphics[width=3.5in]{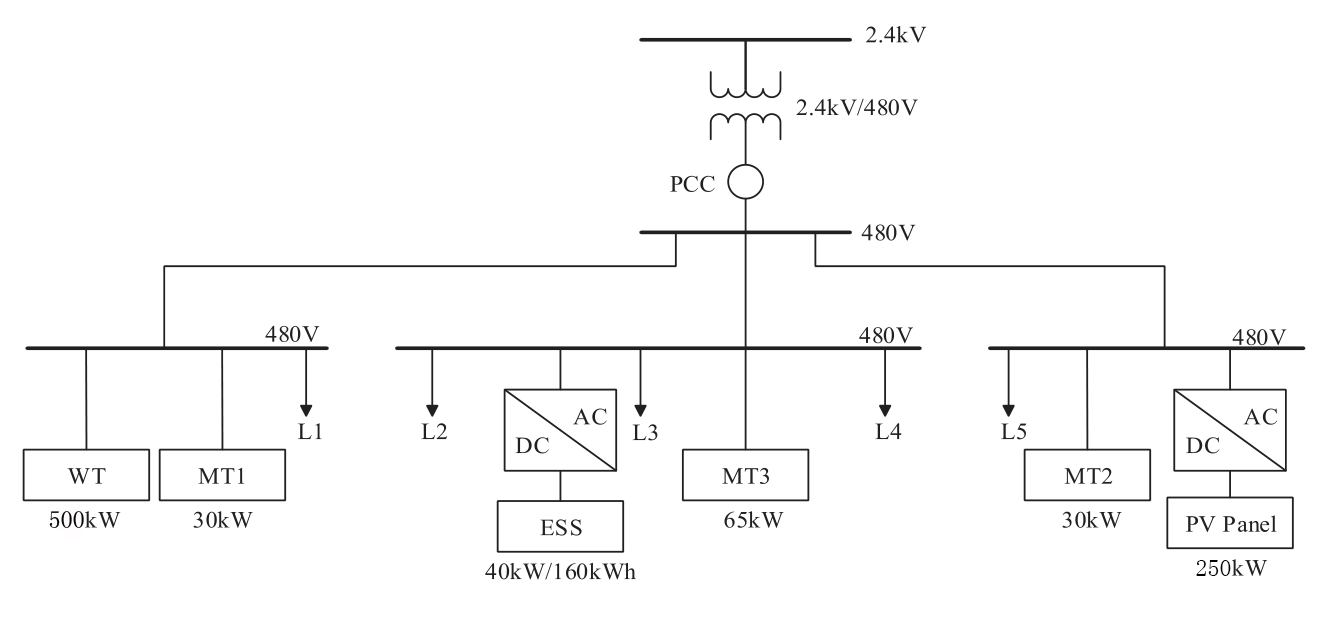}	
	\caption{{\color{black}The microgrid test system.}}	
	\label{fig_microgrid}
\end{figure}

\begin{figure}[t]
	\centering	
	\includegraphics[width=3.7in]{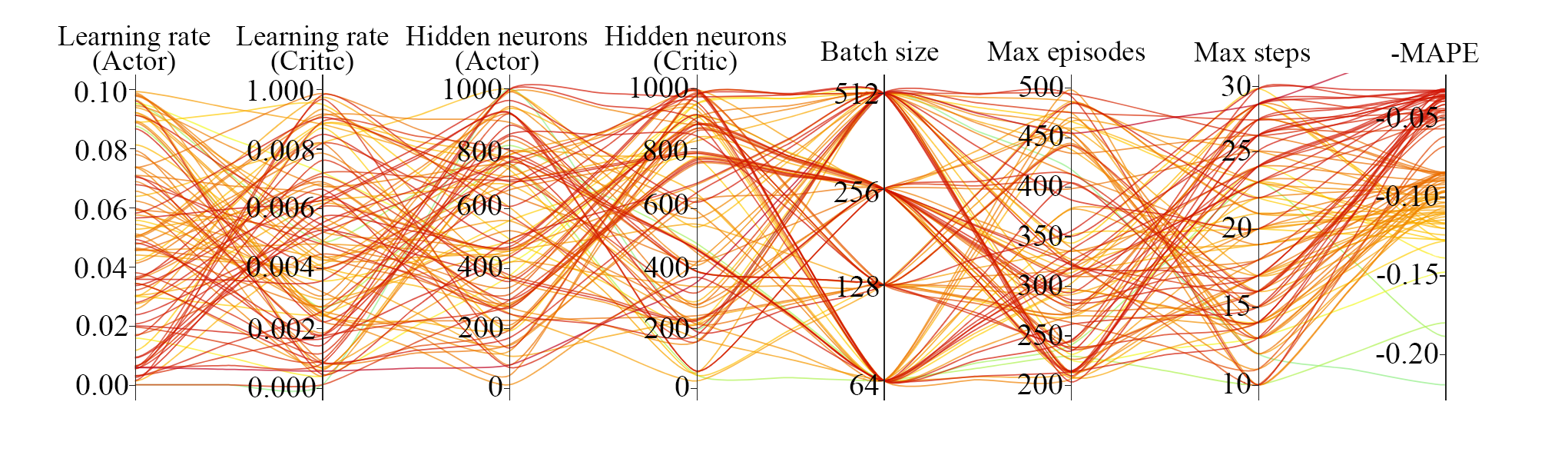}	
	\caption{{\color{black}Hyperparameters optimization.}}	
	\label{fig_hyperparameters optimization}
\end{figure}

\begin{figure}[t]
	\centering	
	\includegraphics[width=3.4in]{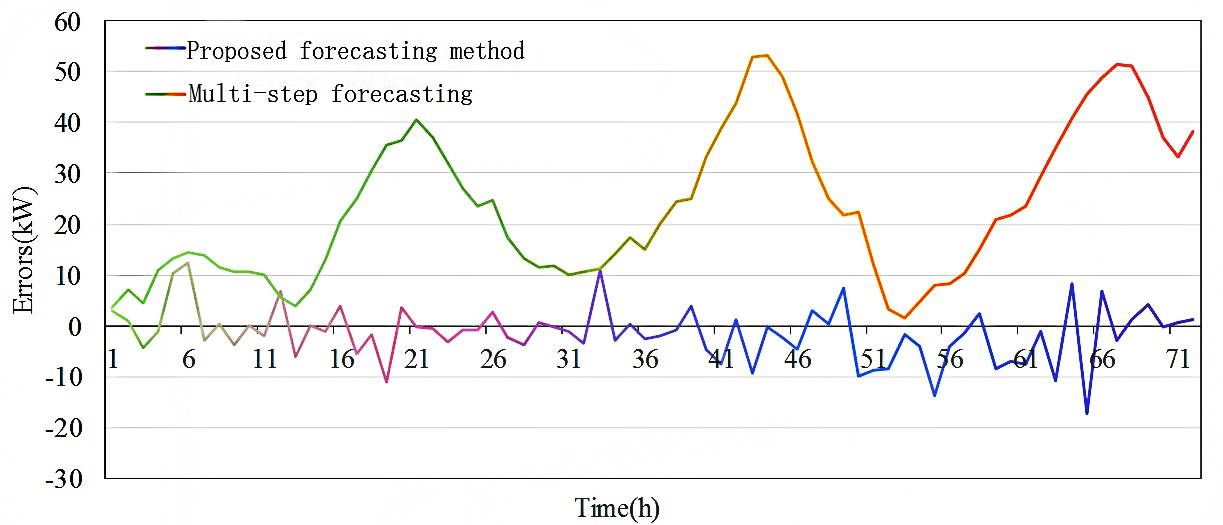}	
	\caption{Forecasting errors with different methods.}	
	\label{fig_error accumulate}
\end{figure}

\begin{figure*}[t]
	\centering	
	\includegraphics[width=7.2in]{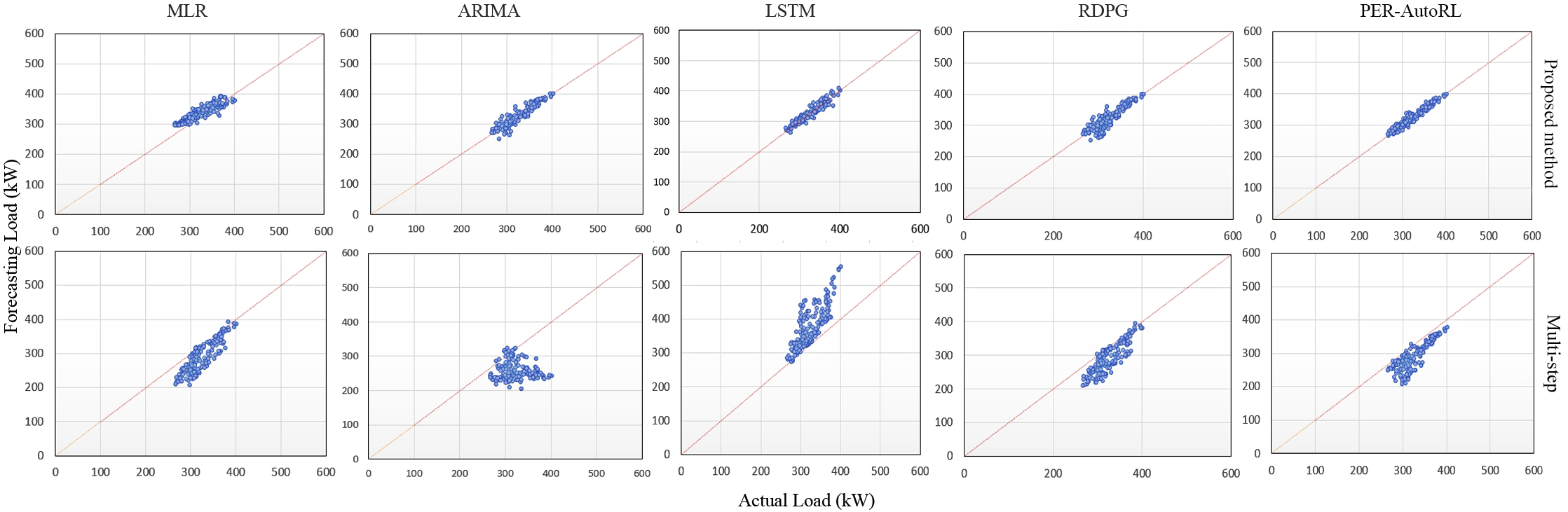}	\caption{{\color{black}Forecasting results with different forecasting models.}}	
	\label{fig_scatter}
\end{figure*}

\section{Case Study}\label{sec.case study}
{\color{black} To examine the effectiveness of the proposed approach, a real-world microgrid in North China is used for numerical simulation analysis.} In this paper, Python is used as the programming language and all simulations are performed on a PC platform with 2 Intel Core dual-core CPUs (2.6Hz) and 6 GB RAM.

\begin{figure}[t]
	\centering	
	\includegraphics[width=3.5in]{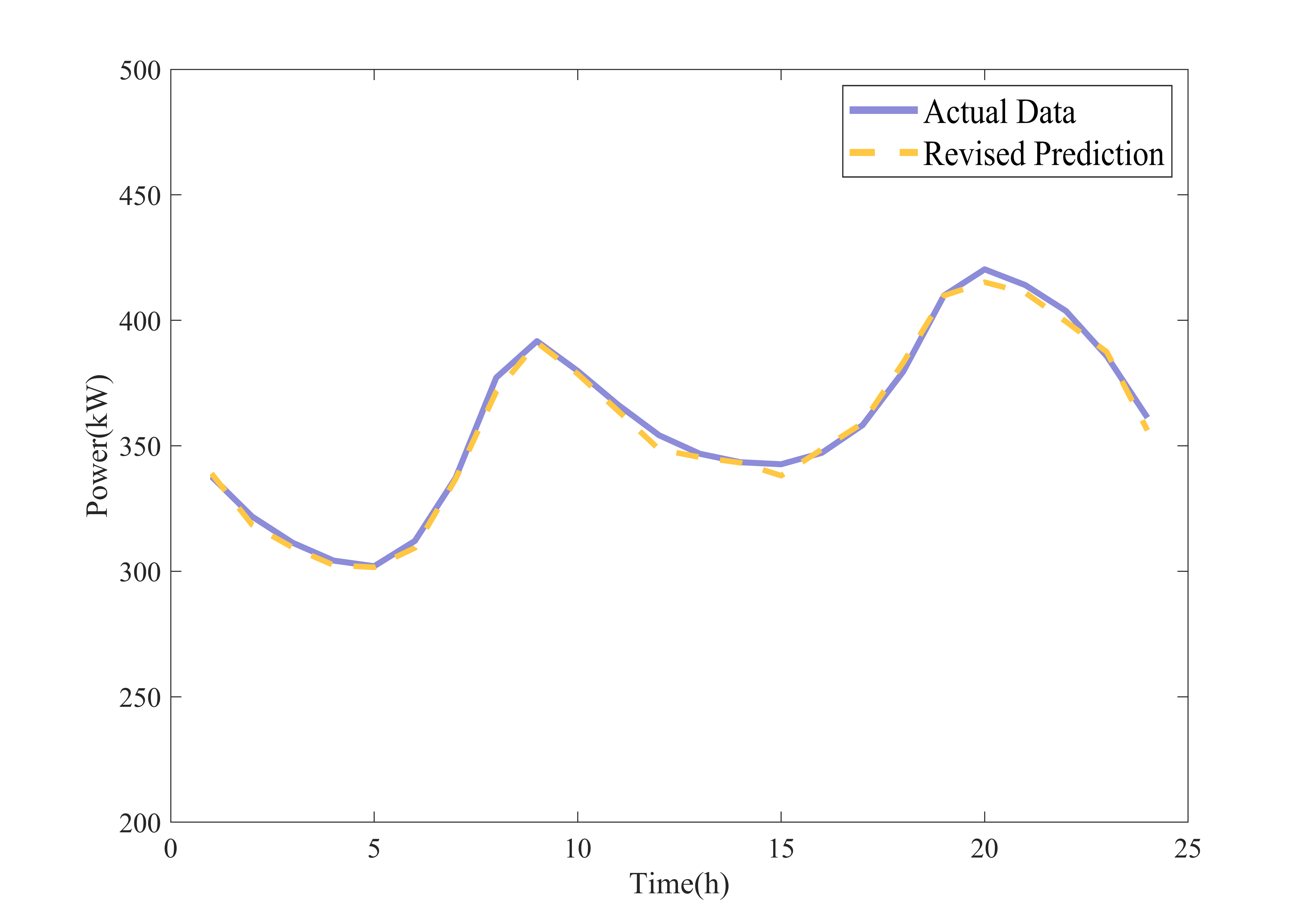}	
	\caption{{\color{black}Forecasting results of loads.}}	
	\label{fig_load}
\end{figure}

\begin{figure}[t]
	\centering	
	\includegraphics[width=3.5in]{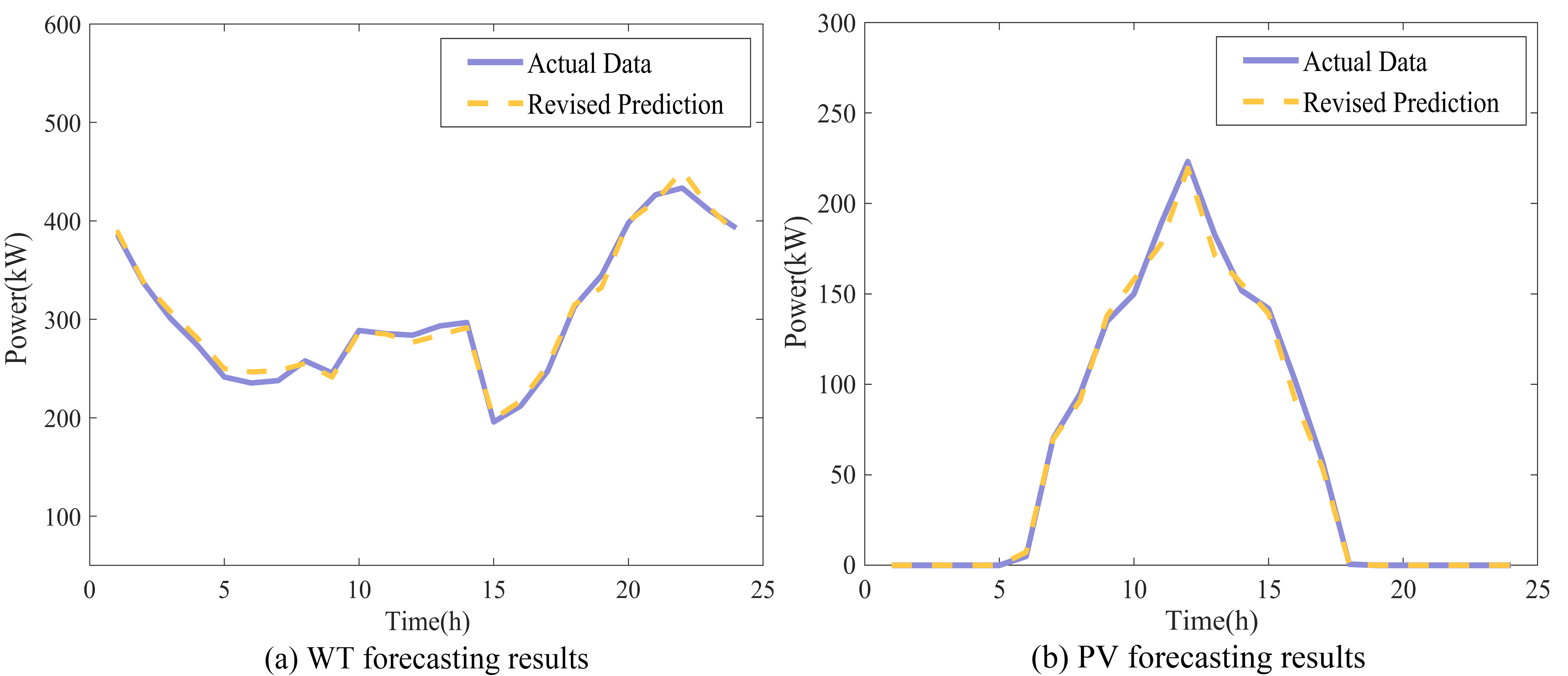}		
	\caption{{\color{black}Forecasting results of WT and PV outputs.}}	
	\label{fig_WP }
\end{figure}

\subsection{ Introduction of Test System}
As shown in Fig. \ref{fig_microgrid}, the system is mainly composed of 3 MT units, a WT unit, a PV panel, and a battery pack. 

\begin{table}[!t]
\caption{ Main parameters of MT units}\label{tab.parameters}
\centering
\renewcommand\arraystretch{1.3}
\begin{tabular}{rclllll}
\hline
\ $\psi(\$)$    &$\xi(\$/Kw)$    &$P_{min}(Kw)$    &$P_{max}(Kw)$  &$\tau(\$)$   &$\zeta(\$)$  &$N$\\
\hline
$1.2$  &$0.35$ &\qquad$5$ &\qquad$30$ &$1.6$ &$0.04$ &\ $2$\\
$1.0$  &$0.26$ &\ \ \quad$10$ &\qquad$65$ &$3.5$ &$0.04$ &\ $1$\\
\hline
\end{tabular}
\end{table}

In order to reasonably verify the rationality of the proposed forecasting method, the 5 years of WT and PV outputs and load data from the microgrid from January 1, 2015 to December 31, 2019 are adopted. Then use the method described in sec. \ref{sec.FM} to divide the datasets into 24 new time series, and regard these as the original datasets for prediction. Finally, the original datasets are divided at a ratio of 0.8:0.2 for model training and testing.

We use this system to verify the feasibility of the scheduling model based on PER-AutoRL forecasting. Tab. \ref{tab.parameters} shows the specific parameters of the MTs in this system.

\par The parameters of lead-acid batteries are as follows: $P_{\max }^{DC}$=$P_{\max }^{DC}=40$kW, $S_{\min}=32$kW$\cdot$h, $S_{\max}=160$kW$\cdot$h, $\eta_{ch}=\eta_{dc}=0.9$.

\subsection{Multi-period Renewable Power Outputs and Load Forecasting}
\par {\color{black}For the prediction problem in this work, the proposed PER-AutoRL will automatically determine the model architecture and hyperparameters based on the input multi-sub time series of WT, PV outputs, and load in a customized manner. Fig. \ref{fig_hyperparameters optimization} shows the optimization results of the hyperparameters required by the PER-AutoRL to model the load data. Each curve in the figure represents a set of hyperparameters, {\color{black}each ordinate axis represents the values of various hyperparameters, and the last ordinate is the negative MAPEs of using these hyperparameter sets}; the darker the color, the more appropriate the hyperparameters. It can be seen from the figure that our designed PER-AutoRL manages to achieve satisfactory optimization results in most cases, and only the MAPEs resulting by three sets of hyperparameters are higher than 15\%. Furthermore, only a few optimization iterations are needed for the PER-AutoRL to find suitable hyperparameters. Therefore, these results prove that our PER-AutoRL is able to automatically determine the most proper model architecture and  hyperparameters in a customized and efficient manner.}

Fig. \ref{fig_error accumulate} shows the load forecasting errors of the microgrid for a total of 72 hours in 3 days. Regarding the multi-step forecasting method, the forecasting results are relatively accurate only in the first few steps; while as time goes by, the error accumulation phenomenon gradually occurs, which leads to poor performance in forecasting. In contrast, when using our forecasting method, the prediction error is stably maintained in a much smaller interval, which confirms that our method is able to handle the error accumulation issue in traditional multi-step forecasting.

\par As illustrated in Fig. \ref{fig_scatter}, the scatter diagram shows the load forecasting results of the proposed method and multi-step forecasting method with different models. The red diagonal line in the figure is ideal fitting line. The closer the fitting of the scatter points is to this diagonal line, the more accurate the prediction results will be. It can be seen from the figure that no matter what kind of prediction models, the results of the proposed prediction method is significantly higher than the multi-step prediction. In the proposed forecasting method, {\color{black}the prediction performance of the PER-AutoRL is significantly better than that of traditional multiple linear regressions (MLR), autoregressive integrated moving average (ARIMA), LSTM, and recurrent deterministic policy gradient (RDPG).}

\par Fig. \ref{fig_load} shows the one-day load forecasting results of the microgrid using the proposed forecasting method. It can be seen from Fig. \ref{fig_load} that the load in this area has a certain regularity. The load gradually increases from 5:00 to 10:00, and decreases during the lunch break from 11:00 to 14:00, then it increases from 15:00 to 20:00. The load gradually reduced again from 21:00 to 5:00. It can be seen that the forecasting results can reflect this trend appropriately.

\par Fig. \ref{fig_WP } shows the forecasting results of WT and PV outputs. The figure reveals that when we adopt the proposed method, no matter the prediction results of WT or PV outputs can fit the real power generation curve suitably.

\par It can be seen from Tab. \ref{tab.MAPE} that, compared with the multi-step prediction method, the forecasting accuracy of the proposed method is significantly improved, {\color{black}and it is further improved by revising the prediction values via the error distribution compared with single-step prediction method,} indicating that the proposed method has achieved the goal of reducing error accumulation and improving prediction accuracy. At the same time, it can be seen that the load forecasting errors are smaller than that of WT and PV outputs, which suggests that the randomness of WT and PV outputs are stronger than that of the load.

\subsection{Optimal Scheduling Based on {\color{black}PER-AutoRL} Forecasting}
\par {\color{black}As a comparison, the renewable energy outputs and load in traditional scheduling methods without forecasting are obtained by discretizing their probability distributions in an expectation form; while the required spinning reserve capacity is determined via the probability distributions of the WT, PV outputs, and load\cite{li2018optimal}. And then, based on the above data as the input data, the scheduling results of the scheduling model without forecasting can be obtained by optimizing the scheduling model.}

\begin{figure}[t!]
	\centering	
	\includegraphics[width=3.5in]{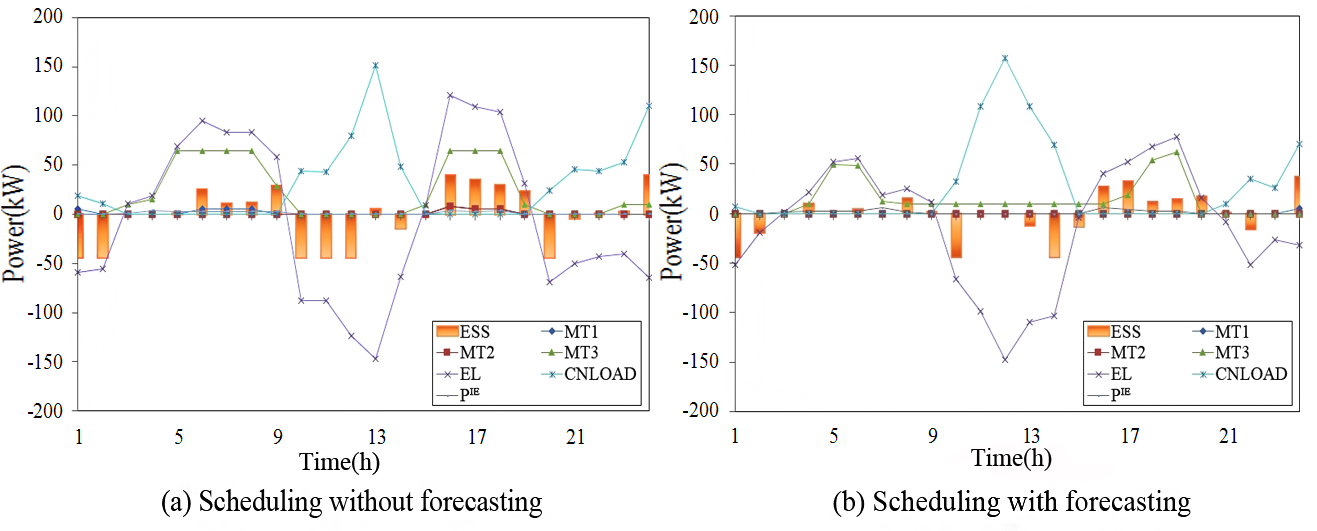} 			
	\caption{{\color{black}Scheduling results.}}
	\label{fig_with without}
\end{figure}

\begin{table}[t]
  \centering
  \caption{{\color{black}Forecasting errors of different methods}}
  \renewcommand\tabcolsep{5.2pt}
       \begin{tabular}{ccccc}
    \toprule
          & Indicators & Proposed  method & Single-step & Multi-step  \\
    \midrule
    \multirow{2}[2]{*}{Load} & MAPE  & {\color{black}0.01825} & {\color{black}0.0215} & {\color{black}0.1416} \\
          & RMSE  & {\color{black}3.9736} & {\color{black}8.8031} & {\color{black}49.9875} \\
\cmidrule{2-5}    \multirow{2}[2]{*}{WT} & MAPE  & {\color{black}0.0709} & {\color{black}0.1015} & {\color{black}0.2227} \\
          & RMSE  & {\color{black}25.8691} & {\color{black}30.5605} & {\color{black}66.6884} \\
\cmidrule{2-5}    \multirow{2}[2]{*}{PV} & MAPE  & {\color{black}0.0813} & {\color{black}0.1261} & {\color{black}0.2611} \\
          & RMSE  & {\color{black}38.8431} & {\color{black}48.7785} & {\color{black}59.8621} \\
    \bottomrule
    \end{tabular}%
  \label{tab.MAPE}%
\end{table}%

\begin{figure}[t]
	\centering	
	\includegraphics[width=3.4in]{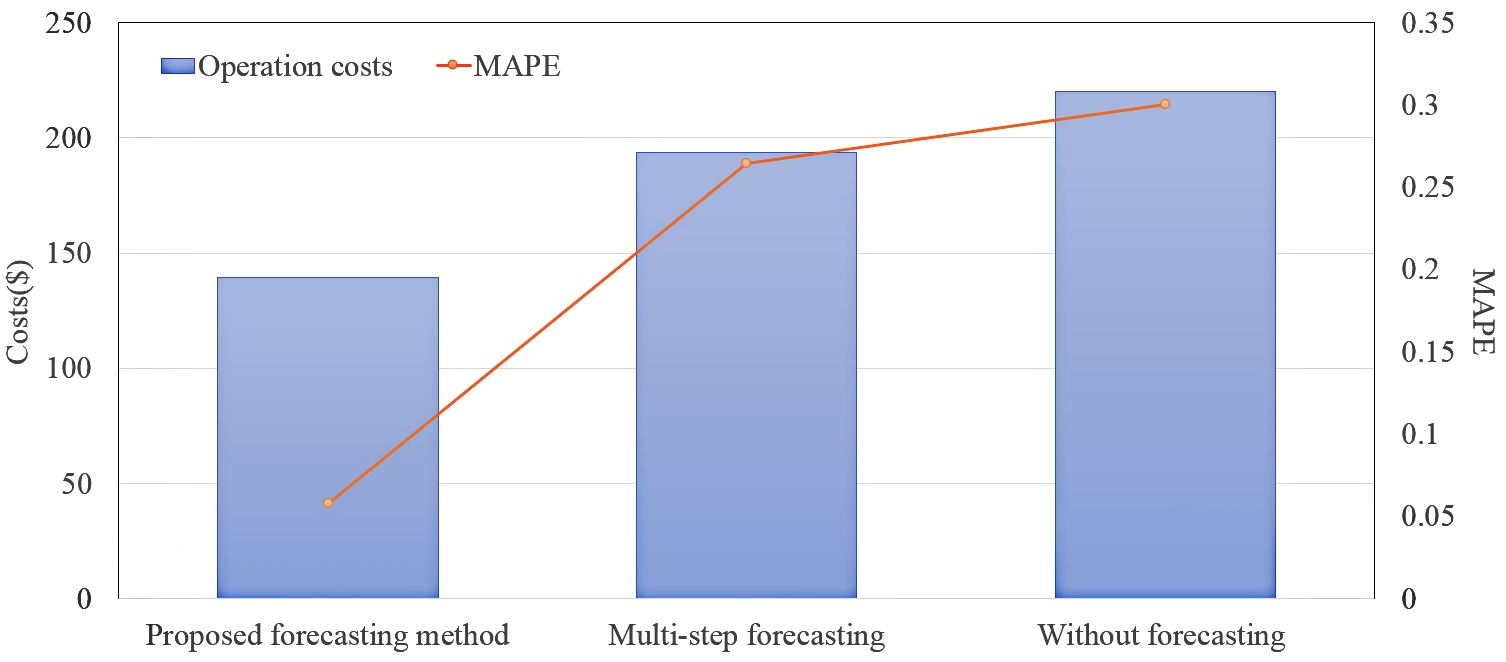} 		
	\caption{{\color{black}Operation costs of scheduling with different prediction methods.}}
	\label{fig_stair}
\end{figure}

\begin{figure}[htb!]
	\centering	
	\includegraphics[width=3.4in]{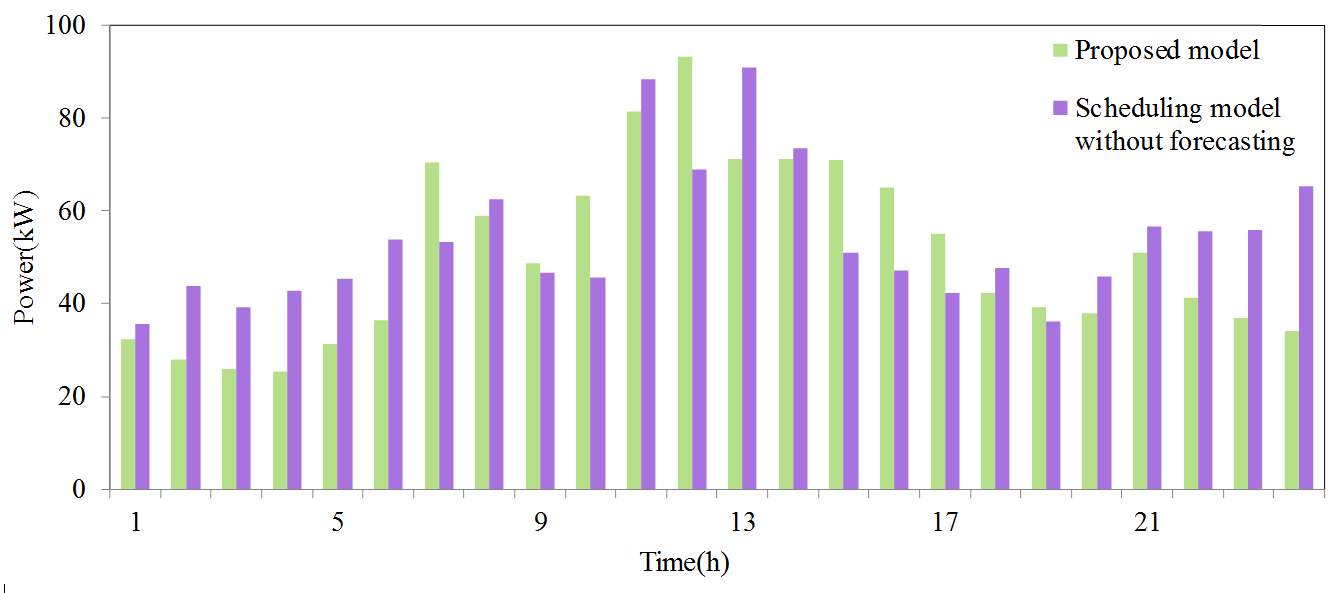} 		
	\caption{{\color{black}Comparison of spinning reserve capacity with traditional scheduling model without forecasting.}}	
	\label{fig_SRC}
\end{figure}

\begin{table}[t!]
  \centering
  \caption{{\color{black}Operating costs at different confidence levels}}
    \begin{tabular}{ccc}
    \toprule
    \multicolumn{1}{c}{\multirowcell{3}{Confidence \\ levels(\%)}} & \multicolumn{2}{c}{Operating cost(\$)} \\
\cmidrule{2-3}          & \multicolumn{1}{c}{\multirowcell{2}{Proposed \\ approach}} & \multicolumn{1}{p{10em}}{Scheduling without forecasting} \\
    \midrule
    90\%  & {\color{black}128.0413} & {\color{black}207.9492} \\
    95\%  & {\color{black}139.3082} & {\color{black}219.9456} \\
    99\%  & {\color{black}206.8478} & {\color{black}253.3075} \\
    \bottomrule
    \end{tabular}%
  \label{tab.costs}%
\end{table}%

\par Fig. \ref{fig_with without} shows the scheduling results with/without a forecasting method at the 95\% confidence level. It can be seen from Fig. \ref{fig_with without} (a) and (b) that the ESS is charged by MTs at the beginning of the scheduling cycle. As shown in Fig. \ref{fig_with without} (a), when the scheduling model without forecasting, MT1 and MT3 provide the spinning reserve to meet the required confidence level and balance the load in most periods. When the load demand increases, MT2 also participates in power generation. {\color{black} As shown in Fig. \ref{fig_with without} (b), the power generated by MT3 and ESS can basically meet the load and spinning reserve demand; it is only necessary to start MT1 at 24:00 and MT2 needn't participate in scheduling.} This fact suggests that our approach is capable of improving the operational economy of the microgrid by reducing the MTs' starting and stopping costs.

\par As shown in Fig. \ref{fig_stair}, the accuracy of the prediction results is closely related to the operating costs, the greater the prediction errors, the higher the operating costs. This fact confirms that different prediction methods have a significant impact on the operating costs of the microgrid.

\par Fig. \ref{fig_SRC} shows the effect of the proposed forecasting method on the required spinning reserve capacity at the 95\% confidence level. Specifically, the spinning reserve capacity required in the proposed scheduling model is less than that of the model without forecasting in 66.67\% periods. This confirms that accuracy forecasting can reduce the spinning reserve capacity significantly.

\par As shown in Tab. \ref{tab.costs}, the spinning reserve capacity cost must be increased with the increase of the confidence level, which will inevitably lead to a increase in the operating costs. Furthermore, Tab. \ref{tab.costs} also illustrates that compared with the scheduling without forecasting, the operating costs of the proposed approach are reduced by {\color{black}18.3\%, 36.7\%, and 38.4\%} at the 99\%, 95\%, and 90\% confidence level, respectively. Therefore, it can be concluded that the scheduling with forecasting is able to effectively improve the operational economy of the microgrid, and that choosing a reasonable confidence level is a key to realize the trade-off between reliability and economy.

\begin{table}[!t]
  \centering
  \caption{{\color{black}Performance comparison}}
    \begin{tabular}{ccccc}
    \toprule
    \multicolumn{1}{c}{\multirowcell{3}{Confidence\\levels(\%)}} & \multicolumn{2}{c}{Proposed approach} & \multicolumn{2}{c}{Traditional approach} \\
\cmidrule{2-5}          & \multicolumn{1}{p{4.7em}}{Operating\newline{}cost(\$)} & \multicolumn{1}{p{4.7em}}{Calculation\newline{}time(s)} & \multicolumn{1}{p{4.7em}}{Operating\newline{}cost(\$)} & \multicolumn{1}{p{4.7em}}{Calculation\newline{}time(s)} \\
\midrule    90\%  & {\color{black}128.0413} & {\color{black}3.7}   & {\color{black}220.8436} & {\color{black}78.2} \\
    95\%  & {\color{black}139.3082} & {\color{black}5.9}   & {\color{black}238.5843} & {\color{black}89.2} \\
    99\%  & {\color{black}206.8478} & {\color{black}5.1}   & {\color{black}249.2354} & {\color{black}118.68} \\
    \bottomrule
    \end{tabular}%
  \label{tab:Performance comparison}%
\end{table}%

\subsection{Performance Comparison}
\par In generally, traditional chance-constrained programming uses Monte Carlo simulations (MCS) to handle the problem, {\color{black}so we use the hybrid intelligent algorithm which combines the particle swarm optimization (PSO) algorithm with MCS in 
\cite{wu2011economic} to solve the chance-constrained model in sec. \ref{sec.model} as a comparison.} In PSO, the population size is 20, and the maximum number of iterations $T$ is set as 150; in MCS, the number of random variables $N$ is 500. Considering that this approach has randomness, the average results of 10 runs are taken. 
\par The results in Tab. \ref{tab:Performance comparison} show that the proposed approach outperforms the traditional approach in calculation efficiency and can significantly reduce the operation costs.

\section{conclusion}\label{sec.conclusion}
How to deal with the uncertainty of renewable energies outputs and load is a crucial issue in an isolated microgrids day-ahead scheduling. {\color{black}In this paper, we for the first time propose a multi-period forecasting method based on PER-AutoRL and integrate it into the microgrid scheduling model considering demand response. The designed PER-AutoRL manages to improve the construction efficiency of the forecasting model in a customized way} and the test results demonstrate that the prediction accuracy of the proposed forecasting method is significantly superior to that of a traditional multi-step forecasting method. Compared with the traditional scheduling methods without forecasting, our proposed approach is able to significantly reduce the operating costs with faster calculation speed.

\par  In future work, we will extend this method to the energy management of integrated energy systems. Besides, exploring advanced technologies to protect the users privacy is also an interesting topic.

\bibliography{./ref}

\bibliographystyle{IEEEtran}

\end{document}